# Boosting Optical Nanocavity Coupling by Retardation Matching to Dark Modes

Rohit Chikkaraddy[1†], Junyang Huang[1], Dean Kos[1], Eoin Elliott[1], Marlous Kamp[1], Chenyang Guo[1], Jeremy J. Baumberg[1]*, Bart de Nijs[1]*

[1] NanoPhotonics Centre, Cavendish Laboratory, Department of Physics, JJ Thompson Avenue, University of Cambridge, Cambridge, CB3 0HE, United Kingdom

**Abstract**

**Plasmonic nano-antennas can focus light to nanometre length-scales providing intense field enhancements. For the tightest optical confinements (0.5-5 nm) achieved in plasmonic gaps, the gap spacing, refractive index, and facet width play a dominant role in determining the optical properties making tuning through antenna shape challenging. We show here that controlling the surrounding refractive index instead allows both efficient frequency tuning and enhanced in/output-coupling through retardation matching as this allows dark modes to become optically active, improving widespread functionalities.**

1. Introduction

Plasmonic nano-constructs with nanometre gaps confine light far below the diffraction limit, with potential applications in single-molecule sensing[1], adatom-catalysis[2], room temperature quantum optics[3–5], and photon harvesting.[2,6,7] Nanoconstructs which incorporate plasmonic nanogaps[8] yield some of the highest[9] and most reproducible[10,11] field enhancements. Strong optical interactions with the metal surfaces slow down light in tightly confined modes, giving effective refractive indices $n_\text{eff} \gg n_g$, dependent on the gap thickness $d$, refractive index $n_g$, and metal permittivity.[8] Inconveniently for applications, the tightest confined modes emit at high angles ($\theta$) to the nanogap normal, leading to poor in/out-coupling.[12] As a result, net optical efficiencies of most nanocavity processes are ripe for enhancement,[21] essential for transitioning nascent technologies into practical applications.

While plasmonic nano-gaps support a few bright nanocavity modes, many modes are dark and only accessible via the near-field.[13–20] Making these bright and accessible at near normal incidence ($\theta$=0), would greatly improve optical access as it provides more operational frequencies and scattering angles, but how to do so is poorly understood and difficult to achieve. Plasmon resonances tune with the surrounding refractive index $n_d$,[21–28] although antenna size, metal, and shape are more commonly employed to tune plasmon resonances instead as these effects have been well characterised and documented. Here, by mapping how $n_d$ enhances specific plasmonic nanocavity mode coupling, we highlight improvements beyond simple wavelength shifts. We attribute this coupling enhancement to improved retardation matching between the slow light of the plasmon and retardation from the high refractive index surrounding medium. Finite-difference time-domain (FDTD) modelling matches comprehensive experimental characterisation of plasmonic nanogap constructs coated in a range of dielectric media of different refractive index. We show how modes shift across the visible, and how dark antisymmetric modes become optically active. These amplified dark modes couple to the far field over a much wider angular range, and critically are experimentally more accessible.

2. Results and discussion

To robustly form identical plasmonic nanogap constructs, a nanoparticle-on-mirror (NPoM) construct is used where a flat Au surface is coated with a molecular self-assembled monolayer (SAM) to form a uniform spacer layer, here biphenyl-4-thiol (BPT) creating a ~1.3 nm thick spacer.[29] Colloidal $D$=80nm

Au nanoparticles (AuNPs) are then deposited on top, forming a NPoM construct of high reproducibility.[11] The optical hotspot in such nanogaps reaches intensity enhancements of $10^6$ and supports a set of optical modes dependent on facet size, shape, polarisation, and gap.[30]

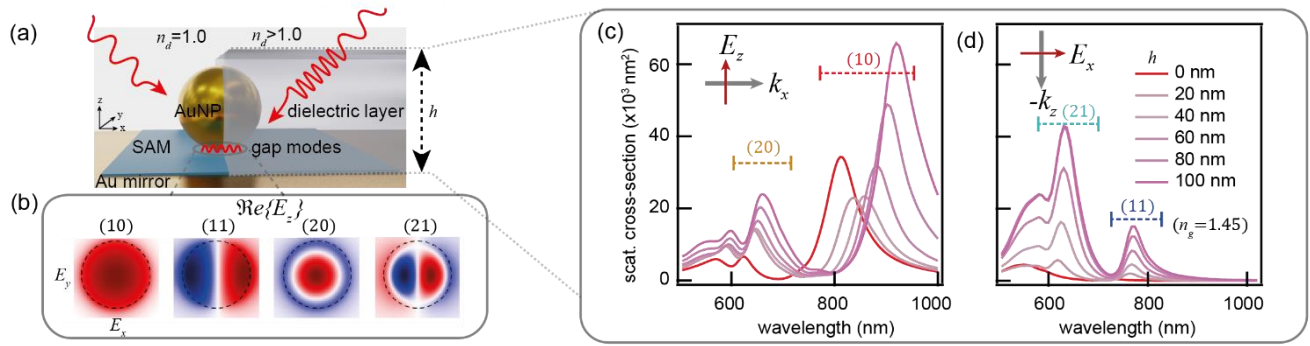

Figure. 1. (a) Left: NPoM geometry in air ($n_d$=1.0) on a Au surface with thin dielectric spacer (1.3 nm, $n_g$=1.45), right: NPoM embedded in $n_d$=1.5 dielectric layer of increasing height $h$. (b) FDTD-simulated near-fields of four lowest energy modes in nanocavity, just above the mirror. (c,d) Effect of dielectric layer height ($h$) on the gap modes under (c) high-angle and (d) normal-incidence illumination.

Full-wave FDTD simulations of these NPoMs truncate the NP to form a 20 nm circular bottom facet, capturing the faceting of colloidal gold nanoparticles (Figure 1a: left).[31] The plasmonic cavity formed between the AuNP and mirror supports a set of optical modes with the four lowest labelled (10,11,20,21).[30] These display characteristic field distributions (Figure 1b), with symmetric 'even' modes (10, 20; denoted as $l0$) and antisymmetric 'odd' modes (11, 21; denoted as $l1$).[30,32] In air ($n_d$=1), the even modes dominantly contribute to the far-field optical properties, whilst the odd modes are non-radiative (dark) and absent from the scattering spectrum. Introducing a high refractive index medium around the metal slows down the incident light, introducing a phase delay between antenna (NP top) and nanocavity (NP bottom), which matches the confined plasmons. Our simulations show that increasing the $n_d$=1.5 dielectric film height ($h$) around the constructs shifts the even modes towards the infrared (Figure 1c) whilst the odd modes steadily become more radiative, as evidenced in scattering intensities (Figure 1d). The scattering strength of (21) mode is comparable to scattering intensities of (20) mode for $h$>80nm, indicating efficient coupling to (21) at normal incidence in contrast with high angle coupling to (20) mode.

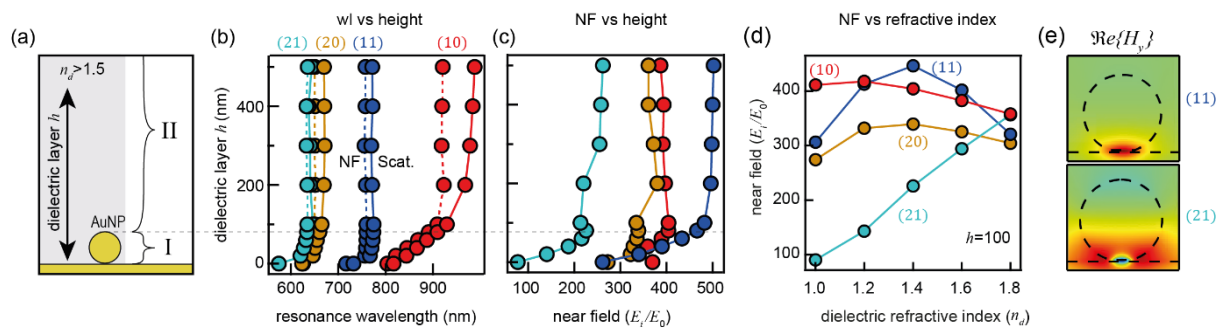

Figure. 2. (a) Scheme depicting two regions of dielectric layer height. (b) Tuning of resonance peak wavelengths extracted from scattering (solid lines) and near-field (dotted lines) spectra for each mode vs $h$. (c) Near-field enhancement ($E_i/E_0$) at spectral peaks of each mode vs $h$. (d) Near field enhancement vs refractive index of embedding dielectric material. (e) Optical field $H_y$ (out of page) around NPoM for odd modes (11, 21) embedded in dielectric coating of $n_d$=1.5, $h$=100 nm.

The nanoparticle's optical properties change most strongly when a film intersects with the spill-out field of the gap and the nearfield of the nanoparticle, Figure 2a region I), and saturates for $h > D$ (region II). This is clearly observed in both near-field and scattering resonance wavelengths (Figure 2b,c). Upon embedding, the nearfield of odd modes is enhanced more than the even modes (Figure 2d), with (21) increasing by ∼250% and (11) by 100% compared to 30% for (20) and 10% for (10) modes (Figure S1a). Using fully embedded geometries ($h$=100 nm) and instead increasing $n_d$ shows the nearfield of the radiative (10) mode decreases (Figure 2d), primarily as its red-shifting resonance is less well confined within the nanogap. The nearfield of the non-radiative (21) mode however greatly increases because of improved in-coupling, becoming comparable to the fundamental (10) mode at $n_d$=1.8. For (11,20) strongest nearfields are observed near $n$=1.4 from the two competing effects. The larger field spill-out of the NP facet for odd modes is visualised from the magnetic field, $H_y$ (Figure 2e). The resonance shifts, increase in scattering intensities and near-field enhancements clearly highlight the importance of refractive index from the surrounding medium in determining the optical properties of plasmonic nanogap constructs.

To evidence these changes experimentally, NPoM geometries are prepared with a range of different refractive index coatings (Figure 3). Dielectric layers 100 nm thick with refractive index $n_d$=1.49, 1.59 or 1.78 are spin-coated onto the NPoMs described above. Averaged dark-field (DF) spectra (Figure 3a) of many hundred NPoMs show how the plasmonic modes evolve with increasing refractive index. The scattering intensity from polymer-coated NPoM nano-antennas is over three-fold brighter than NPoMs in air ($n$=1), due to the improved in-/out-coupling of light. Upon coating, the dominant (10) mode visible at 810 nm in air disappears (red shifting out of the detection range) and higher-order modes red shift and increase in intensity. Extracting the dominant peak position for each refractive index (Figure 3b) shows higher-order modes at 650 nm, 695 nm, 710 nm for $n$=1.49, 1.59, 1.78 respectively.

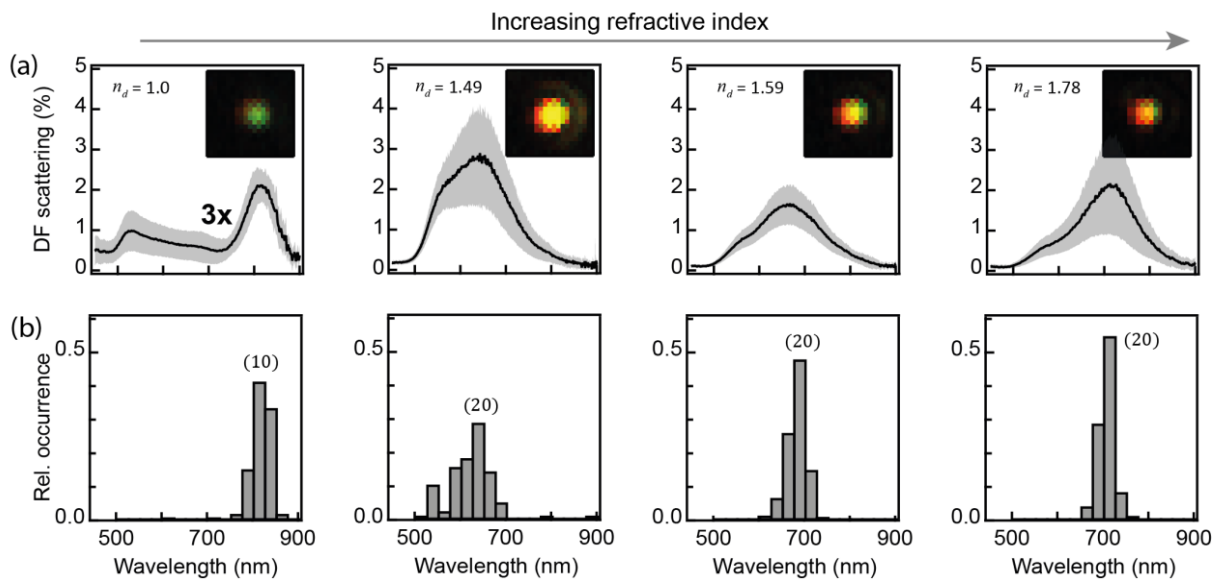

Figure 3. (a) Experimental dark-field (DF) scattering spectra for NPoMs ($D$=80 nm, 1.3 nm spacer) inside progressively higher refractive index coatings ($n_d$= 1.0, 1.49, 1.59, 1.78), note $n_d$=1.0 multiplied 3x for visibility. Black line is average of 1550, 313, 438, 2235 NPoMs respectively, 50% confidence interval in grey. Insets: average DF scattering images. (b) Relative occurrence of main DF visible spectral peak, which red-shifts with increasing refractive index. The (10) mode at $n_d$=1.0 redshifts outside the detection range (>900 nm) for $n_d$≥1.2.

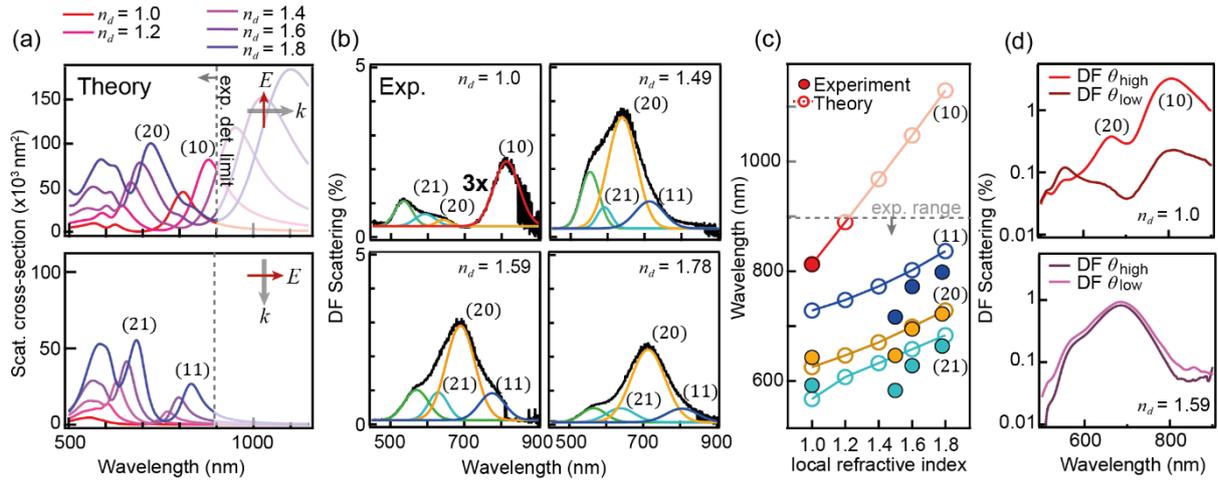

Figure 4. (a) Calculated DF scattering spectra in dielectric media of refractive indices $n_d$ = 1.0-1.8 showing increasing redshifts. Simulations use two different illumination conditions: (top) high-angle with $E \perp$ mirror surface, and (bottom) normal illumination. Insets show $E$, $k$ directions. (b) Experimental DF scattering spectra for $n_d$=1.0, 1.49, 1.59, 1.78 using unpolarised illumination at high angles. Spectra separated into optical modes using multi-Gaussian fit, assigned to different modes. (c) Extracted (solid) and modelled (open) peak positions *vs* surrounding refractive index. (d) Angle-resolved DF scattering spectroscopy of NPoMs shows high angles dominate for $n_d$=1.0 (top), but low angles dominate for $n_d$=1.59 (bottom).

Modelling the effect of refractive index on the fully embedded NPoMs ($h$>100 nm) reproduces the redshifts and rise in scattering intensity of all modes with increasing $n_d$ (Figure 4a). Comparing the simulations with experimental DF spectra (Figure 4b) enables assignment of the dominant modes, (10): red, (20): yellow, with the satellite peaks tentatively assigned to (11,21). The peak positions of these modes are in agreement with predictions (Figure 4c), except for $n_d$=1.49 where all peaks are blue shifted (possibly due to coating morphology under NPs). We note that simulations here also do not capture variations in nanoparticle facet shape, which further breaks the degeneracy of $(l1)$ modes.[30,33]

The simulations predict a significant increase in scattering intensity from the initially-dark odd modes, which emerge as satellite peaks in the DF spectra (Figure 4b,c). The modes can be distinguished by characterising their different out-coupling angles. Even modes (with vertical dipoles) should emit at high angles, and dominate radiation for $n_d$=1.0 when separating high-angle (emitted flux at $\theta_{\text{high}}$= 55-64°) from low-angle ($\theta_{\text{low}}$= 0-10°) scattering in $k$-space spectroscopy (Figure 4d:top, Figure S3a).[34] In contrast, for $n_d$=1.59, nearly equal radiant intensities are simulated for low and high collection angles (Figure 4d:bottom, Figure S3b). This confirms that out-coupling from high-index coated NPoMs is at lower angles, yielding high collection efficiencies even in low numerical aperture systems.

Modelling the scattering from different incident angles (Figure 5) shows that even NPoM modes (which dominate for $n$=1.0) only accept incident light above 45°, whereas odd modes couple to incident light at angles from 0-60° (Figure 5b). Increasing the efficiency of the latter modes is thus critical as most incident light arrives at angles <45°, even for a high NA illumination, as illustrated for a collimated Gaussian beam over-filling the back-aperture of a 0.9 NA objective (Figure 5a, 'laser irradiation'). The angular scattering of DF light from NPoMs is experimentally measured using $k$-space imaging on $n_d$=1.0 and 1.59 samples (Figure 5c, see supporting information note 3, Figure S3 for details). At $n_d$=1 NPoMs scatter near 60° as predicted, while $n_d$=1.59 coated NPoMs scatter over a wide angular range between 0-55°. This confirms that odd modes dominate emission when NPoMs are embedded in higher refractive index surroundings.

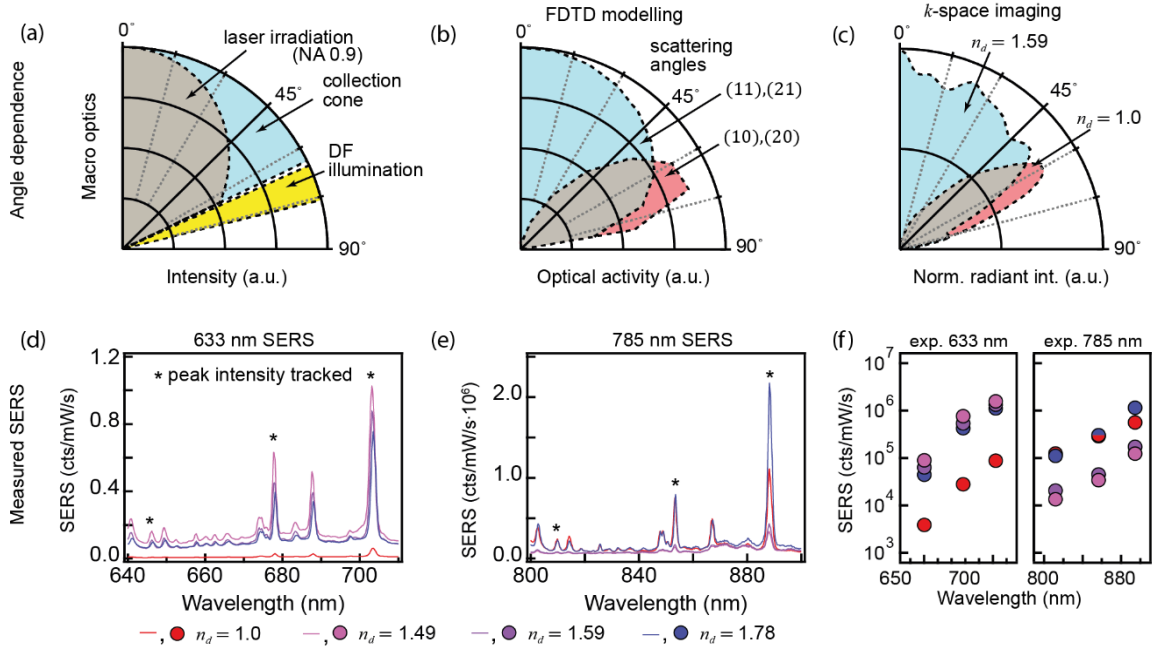

Figure 5. (a) Experimental numerical apertures (NA) for DF illumination, collection, and laser irradiation cones. (b) Calculated NPoM excitation and radiation cones for both even and odd modes. (c) Experimental $k$-space scattering from NPoM with ($n_d$=1.59, blue) and without ($n_d$=1.0, red) dielectric coating. (d,e) SERS spectra for NPoMs in air or dielectric coatings from $n_d$=1.49-1.78, as well as colloidally-grown individual nanolenses ($n_d$=1.49), using (d) 633nm and (e) 785nm lasers. (f) Extracted peak intensities (peaks ∗ in d,e) for each NPoM geometry at 633nm/785nm pumping showing improved performance.

These benefits from retardation matching can provide significant performance improvements in nonlinear-processes from plasmonic nanogap constructs. To demonstrate this, surface-enhanced Raman spectroscopy (SERS) spectra are recorded from the BPT gap molecule using 633 nm and 785 nm lasers for each refractive index, normalised to the laser power, and corrected for the instrument response[35] (Figure 5d,e). For 633 nm, embedding gives up to 20x SERS increase, with comparable performance for refractive indices $n_d$=1.49, 1.59, 1.78. Apart from this enhancement, the SERS spectra are nearly identical showing no additional signal components from the embedding dielectric film (since the SERS originates from the strong hotspot inside the gap, Figure 1a).

Extracting amplitudes of 3 SERS peaks (∗) for each refractive index isolates the vibrational signals from changes in background and noise, and demonstrates the clear enhancements from embedding at every spectral position (Figure 5f). For $n_d$=1, the highest SERS signals are collected for 785 nm excitation, as expected from the strong (10) mode at 810 nm. However, when the surrounding refractive index is increased to $n_d$=1.49, 633 nm SERS signals increase by >12x but SERS signals from 785 nm excitation drop since the (10) mode shifts out of resonance for 1.4< $n_d$ <1.6. The latter SERS intensity recovers for $n_d$=1.78, when higher order modes shift into resonance, with 633 nm SERS further increasing to 23x (Figure 5f).

To gain a better insight into these enhancements and distinguish them from wavelength tuning, a simple nanocavity model is devised. Simulations of scattering spectra and near-field enhancements give the parameters $E^2, V, Q$ for the near-field intensity enhancement, mode volume, and quality factor of each mode (Supporting information note 4). The mode coupling efficiency $C_{lm}$ into the nanocavity is then

$$C_{lm} = \frac{E^2 V n_g^2}{8\pi Q R^3} \quad (1)$$

This extracted coupling rate for the (10) mode increases by 50% as the NPoM is progressively covered with an $n_d$=1.6 coating (SI). The coupling of the higher order odd modes (11),(21) increases by >200% (see SI). This again shows that the initially dark modes become much brighter, through the increased retardation of light through the dielectric layer. The effective optical path between the NP top and bottom approaches $\lambda/2$, which then matches the magnetic-type coupling of ($l1$) as clearly seen by the phase difference across the NP in Figure 2e for the (21) mode.

Whilst specific to this geometry, similar enhancements are expected for other embedded nanocavities. A further valuable feature of the polymer coating is that it increases nanoarchitecture lifetimes. The optical properties of such NPoMs are preserved over more than a year when covered with a polymer, whereas NPoM geometries exposed to air degrade within a month. This improvement in chemical stability is likely due to oxygen and moisture exclusion.

## 3. Conclusions

In summary, we show through both simulation and experiment how tuning of the surrounding refractive index can improve the optical performance of plasmonic nanocavity constructs. Increases in refractive index dramatically improve the acceptance and out-scattering angles of such structures. Experiments and simulations show this occurs by increasing the optical coupling of the antisymmetric odd modes in the plasmonic nanogaps, and gives more than ten-fold amplification in signal intensities for the same incident laser power in SERS sensing applications. These results are more generally applicable to a wide range of nanogap plasmonic structures, since the field orientations perpendicular to the metal surfaces are universal, although the details of peak positions and angles will vary for individual constructs as will the antenna mode coupling. This understanding should encourage new strategies to further boost the optical performance of plasmonic nanostructures.

4. Methods

*4.1. Sample preparation:* All chemicals were ordered from Sigma Aldrich, unless noted otherwise, and used as received. To make the NPoM geometry an atomically flat (111) silicon wafer was coated with a 100 nm Au film using a Lesker E-beam evaporator at a rate of 0.1 Å/s. Then 2 μL droplets of a 2-part epoxy glue (Epo-Tek 377) were deposited on the Au coated wafer to attach Si chips of size approximately 5 mm. The epoxy was cured at 150°C for 2 hours and the wafers were gradually cooled back down to room temperature. The Si chips were peeled off the wafer, exposing a clean, flat Au surface which was transferred to a 1 mM solution of biphenyl-4-thiol (BPT; 97%) in ethanol (≥99.5%, absolute) and left overnight. The BPT coated samples were rinsed with ethanol and blown dry using nitrogen, and 80 nm AuNPs were deposited by resting a 10 μL drop of colloid suspension (BBI Solutions, OD1, citrate stabilised) on each of the samples for 20 seconds.

Dielectric layers were deposited by spin coating a polymer onto NPoM samples. Thickness and refractive index were measured on reference Si samples for each coating type with a spectroscopic ellipsometer (Alpha-SE, J. E. Woollam) using a Cauchy fit for the refractive index. Solution concentration and spin coating speed were tuned to obtain a final film thickness close to 100nm. The quoted nominal refractive index is the value at 650 nm.

Index $n_d = 1.49$: polymethyl methacrylate dissolved in anisole at 2 wt% (commercial PMMA A2 solution by MicroChem). Spun at 1500 rpm with 500 rpm/s acceleration. Film thickness 112 nm. Cauchy coefficients (MicroChem datasheet): A=1.478, B=7.204·10$^{-4}$, C=-3.478·10$^{-4}$.

Index $n_d = 1.59$: Poly(2-chlorostyrene) dissolved in chloroform at 1wt%. Spun at 4000rpm with 500rpm/s acceleration. Film thickness 119 nm. Cauchy coefficients: A=1.588, B=3.19·10$^{-3}$, C=8.2·10$^{-4}$.

Index $n_d = 1.79$: Poly(pentabromophenyl methacrylate) dissolved in anisole at 7 wt%. Spun at 2000 rpm with 500 rpm/s acceleration. Film thickness 101 nm. Cauchy coefficients: A=1.779, B=-4.45·10$^{-3}$, C=4.79·10$^{-3}$.

*4.2. NPoM characterisation:* DF spectra were collected using an Olympus BX51 microscope, fibre coupled to an Ocean insight QE65Pro spectrometer. In house particle tracking software was used to identify and characterise individual NPoM geometries, see ref[36] for details. For SERS, spectra were collected using a homebuilt Raman spectroscopy setup consisting of two single frequency diode lasers (633 nm and 785 nm), a Triax 320 spectrometer with a 150 l/mm grating paired with a back illuminated EMCCD. The relative low lines/mm grating allows for simultaneous collection of SERS at 633 and 785 nm excitation.

*4.3. K-space imaging, and Angle-resolved DF scattering spectroscopy:* Individual nanostructures are illuminated with focused incoherent white light at an annular illumination angle of 64–75° with respect to normal incidence. Scattered light at <64° is collected through a DF objective (Olympus 100xBD, NA 0.9). The scattering pattern is determined using the light intensity distribution in the back focal plane of the microscope objective. Single nanostructures are spatially isolated by spatially filtering the magnified real image plane with a pinhole. The back focal plane image is demagnified by 3 times before being imaged on the entrance slit (150 µm wide) of a Triax 320 spectrometer, where a narrow range of the scattering pattern near $k_x/k_0 = 0$ is filtered and dispersed by a grating and collected using an Andor Newton 970 BVF EMCCD (Figure S3). Using an MFP-3D AFM System (Asylum/Oxford Instruments) the flatness of the polymer coated NPoMs was characterised (Figure S8), yielding an overall standard deviation in height of 2.7 nm and showing occasional small (2-9 nm) bumps, which we attribute to the covering of nanoparticles, and larger bumps likely arising from air bubbles or dust. Measuring between 'large' bumps the standard deviation drops to 0.65 nm.

*4.4. FDTD simulations*: Full-wave 3D simulations are performed using Lumerical FDTD Solutions. The Au NP is modelled as a truncated sphere (with facet width of 20 nm) of radius 40 nm on top of an infinite dielectric sheet of refractive index of $n_g = 1.45$ and gap size of 1.3 nm matching the BPT thickness.[29] The thickness of the Au slab placed below the BPT layer is infinite to the perfectly matching layer (PML) and the AuNP is embedded into a dielectric film of different heights and refractive index ($n_d$) as mentioned in the text. The NPoM geometry is illuminated with a plane-wave with polarization either perpendicular or parallel to the metal surface to access different sets of modes.

Supporting information
Source data can be found at DOI link: https:// (to be provided on acceptance).

**Author information**
Corresponding Authors
* Dr Bart de Nijs, bd355@cam.ac.uk;
* Prof Jeremy J Baumberg, jjb12@cam.ac.uk;
Authors
† Dr Rohit Chikkaraddy, present address: School of Physics and Astronomy, University of Birmingham, Edgbaston, Birmingham B15 2TT, UK

Funding

The work is supported by the European Research Council (ERC) under Horizon 2020 research and innovation programme PICOFORCE (Grant Agreement No. 883703), THOR (Grant Agreement No. 829067) and POSEIDON (Grant Agreement No. 861950). JJB acknowledges funding from the EPSRC (Cambridge NanoDTC EP/L015978/1, EP/L027151/1, EP/S022953/1). R.C. acknowledges support from Trinity College, University of Cambridge. BdN acknowledges support from the Royal Society (URF\R1\211162), BdN and MK acknowledge support from the Winton foundation for the Physics of Sustainability. MK acknowledges support from EPSRC Airguide Photonics grant EP/P030181/1.

Notes

The authors declare no competing interest.

Supporting Information

Boosting Optical Nanocavity Coupling
by Retardation Matching to Dark Modes

Rohit Chikkaraddy[1], Junyang Huang[1], Dean Kos[1], Eoin Elliott[1], Marlous Kamp[1], Chenyang Guo[1], Jeremy J. Baumberg[1]*, Bart de Nijs[1]*

[1] NanoPhotonics Centre, Cavendish Laboratory, Department of Physics, JJ Thompson Avenue, University of Cambridge, Cambridge, CB3 0HE, United Kingdom

# Supporting information

Supplementary note 1: **Nanogap mode nearfield enhancements with increasing surrounding refractive index.**

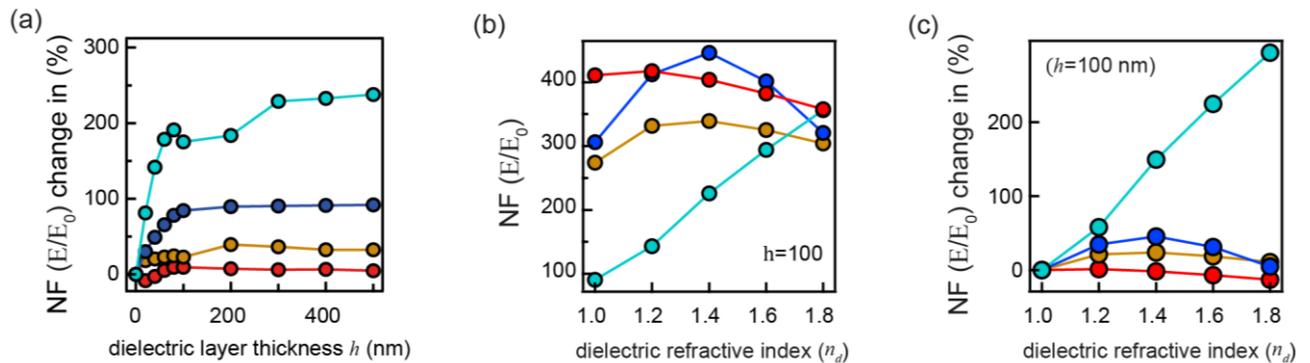

**Figure S1: Effect of surrounding refractive index on the nearfield of plasmonic gap modes.** a) Percentage enhancement of each mode with increase in $h$ for a high refractive index ($n_d$=1.5) dielectric coating. b) Nearfield enhancement as a function of the refractive index of the embedding dielectric medium ($n_d$). c) Relative nearfield enhancements in percentage vs refractive index of the embedding dielectric medium ($n_d$).

Supplementary note 2: **Illumination geometries**

To determine the illumination geometry of the objective used, the DF illumination was visualised on a white sheet of paper placed perpendicular to the image plane. The angle of DF illumination was measured to be 75° (Figure S2).

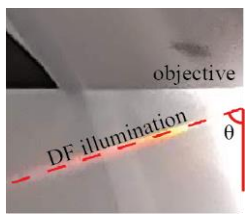

**Figure S2: Darkfield illumination angle (θ).** Measured for an MPLFLN-BD 0.9NA 100x Olympus objective, visualized on a white piece of paper placed perpendicular to the focal plane.

Supplementary note 3: **Angle dependent scattering spectroscopy**

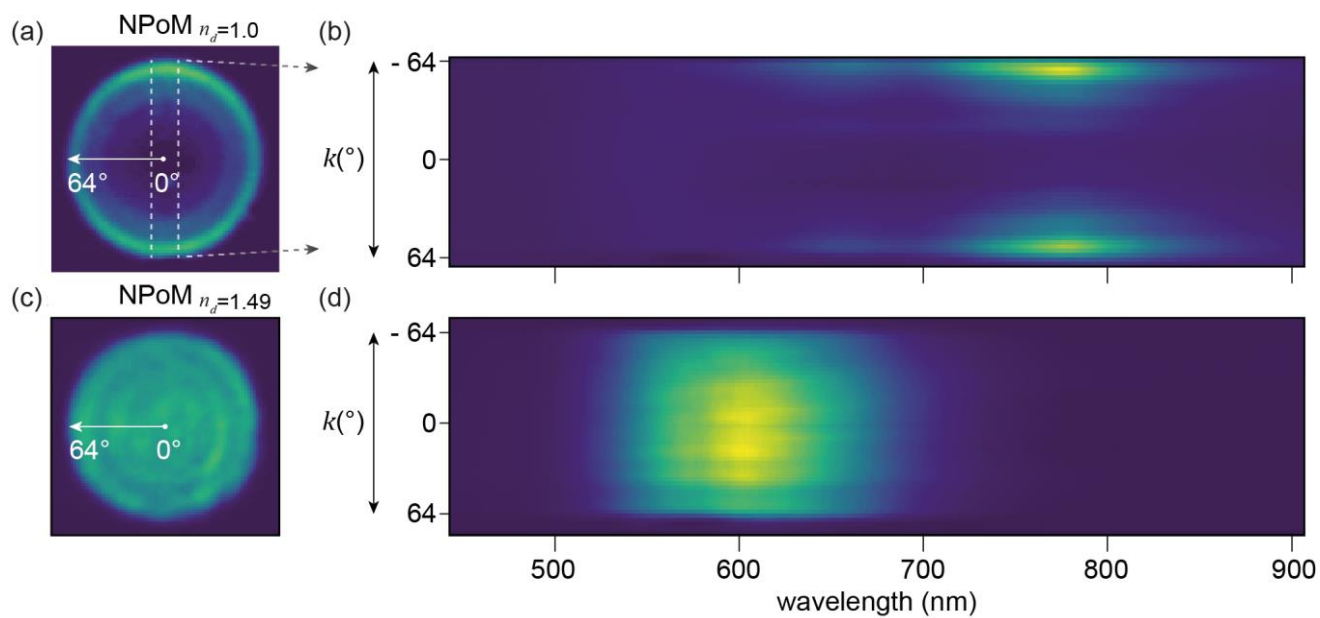

**Figure S3:** Back focal plane (BFP) dark field scattering images captured from a single NPoM (a) with and (c) without refractive index coating ($n_d$=1.0, $n_d$=1.49 respectively) using a NA=0.9 dark field objective. (b,d) Corresponding angle-resolved DF scattering spectra captured by filtering a wavevector range of the BFP image near $k_x/k_0$=0 (dashed lines in a) and dispersed by a 150 lines/mm grating.

Supplementary note 4: **Analysis of Scattering and Near-field in NPoMs**

FDTD simulations give the spectrum of scattering $S(\lambda)$ and the near-field enhancement $\mathrm{E}(\lambda) = E_i/E_0$. From this the resonant peak $\lambda_n$ can be extracted for each mode ($n = 10, 20, 11, …$) which is similar in both $S$ and E. From FDTD, then 3 parameters can be extracted $\{S_{pk}, \mathrm{E}_{pk}, \Delta\lambda\}$ for each mode (Figure S4).

Scattering $S$ (in units m²) comes from scattered power (in W) = $S$ x incident intensity (W/m²). The scattered power per cycle (period $\lambda/c$) is the total energy in the cavity ($U$) x fraction radiated per cycle ($l_r$). The energy in the cavity is $U = V\frac{1}{2}\varepsilon_g\varepsilon_0 E_i^2$ for cavity volume $V$ for each mode, while the incident intensity is $\frac{1}{2}c\varepsilon_0 E_0^2$. Hence

$$U l_r \frac{c}{\lambda} = S \frac{1}{2} c \varepsilon_0 E_0^2$$

or rearranging

$$l_r = \frac{S}{\mathrm{E}^2} \frac{\lambda}{V \varepsilon_g}$$

which is dimensionless as required. We use the peak values on resonance for $S, \mathrm{E}$.

To obtain the mode volumes, the nearfield profile across the gap is used, $\mathrm{E}(\lambda_n, x)$. The mode volume is defined as

$$V = \frac{\int \mathrm{E}^2 dV}{\max(E^2)} = \frac{2\pi d \int \mathrm{E}^2 r dr}{\max(E^2)}$$

with an extra factor of ½ for the odd (11),( 21) modes which have an additional sin(φ) dependence. Note that the volume so far omits out the field which penetrates slightly into the metal facet on either side of the nanocavity, as well as the much weaker field around the rest of the NP. The decay length $\delta$ of light into the metal follows from the resonant wavevector $k_\| w = 2\alpha$ for facet diameter $w$ and Bessel zero $\alpha$, since $k_\perp \sim i k_\|$ and $\delta = k_\perp^{-1} = w/2\alpha$ ~5 nm, thus larger than $d$. The extra metallic volume contribution is then $2.\delta.\pi\left(\frac{w^2}{4}\right)$ but the field is smaller by $\frac{\varepsilon_m}{\varepsilon_g} = \frac{w}{d\alpha}$, which gives $V_m = \varepsilon_m \left(\frac{d\alpha}{w}\right)^2 \frac{w^3\pi}{4\alpha} = \varepsilon_g w^2 d\pi/4$ ~ 850 nm³, which is twice as large as the original gap volume. As a result, the total volume is three times the original integration above. Indeed to match previous estimates for $l_r$~0.6 at $h$=0 (from COMSOL calculations), a volume roughly double the integration volume is required.

Note that the integrated scattered flux $S$ is derived from the peak scattered flux $S_{pk}$ (in a particular direction). This is different for the normally radiating modes ($n1$) and the high angle modes ($n0$), and is evaluated using calculated mode patterns as shown in ref[1].

The total Q-factor can be extracted from the linewiths, $Q_t = \lambda_n/\Delta\lambda_n$, but we note that there is considerable confusion in the literature about whether an extra prefactor of $2\pi$ is included. To reproduce the known values of $l_r = 0.7$ at $h$=0, the appropriate formula is $Q_t = 2\lambda_n/\Delta\lambda_n$. Since $Q = 2\pi/l$, we extract $l_t = 2\pi/Q_t$ and
$$l_{nr} = l_t - l_r$$
so that the radiative efficiency for each mode can be defined as
$$\eta_S = l_r/(l_r + l_{nr})$$

Finally, we obtain the field enhancement inside the nanogap. In ref[2], the antenna mode coupled energy was estimated from the polarizability of the NP to give the energy coupled into the nanogap
$$\tfrac{1}{2}\varepsilon_0\varepsilon_g E_i^2\ V = Q_t.2\pi\varepsilon_0 R^3 \beta E_0^2$$
where $\beta$~2 is the spectral enhancement here, which gives
$$E^2 = \frac{E_i^2}{E_0^2} = \zeta\frac{Q_t}{V}\frac{8\pi}{\varepsilon_g}R^3$$
with an extra coupling factor $\zeta$ included for how well each mode couples, and which describes the effective antenna cross section.

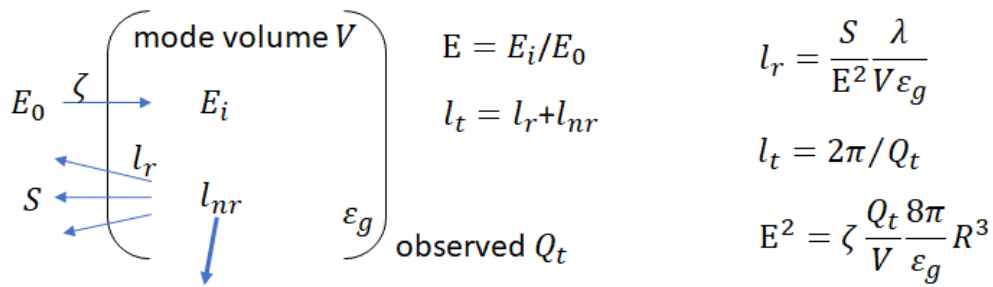

**Figure S3: Summary of model and definitions**

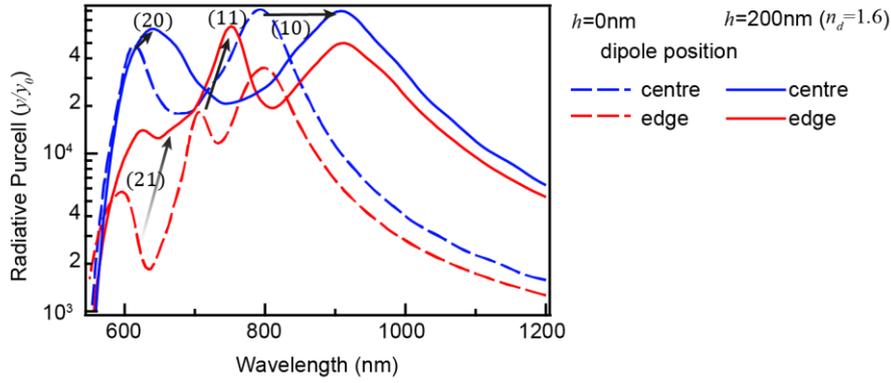

**Figure S4: Radiative Purcell** measured at the centre and edge of the nanocavity for both for no dielectric layer and a layer height of $h=200$nm, from which the change in scattering intensity is extracted for each mode.

As noted above, we scale $V$ and $Q$ to match the calculated $l_{r,nr}(h=0)$ for (10) for coating refractive index $n_d = 1.6$. The extracted parameters for each mode are plotted below (Figure S5), except for (21) as the initial uncoated intensity of this mode is too weak to be extracted.

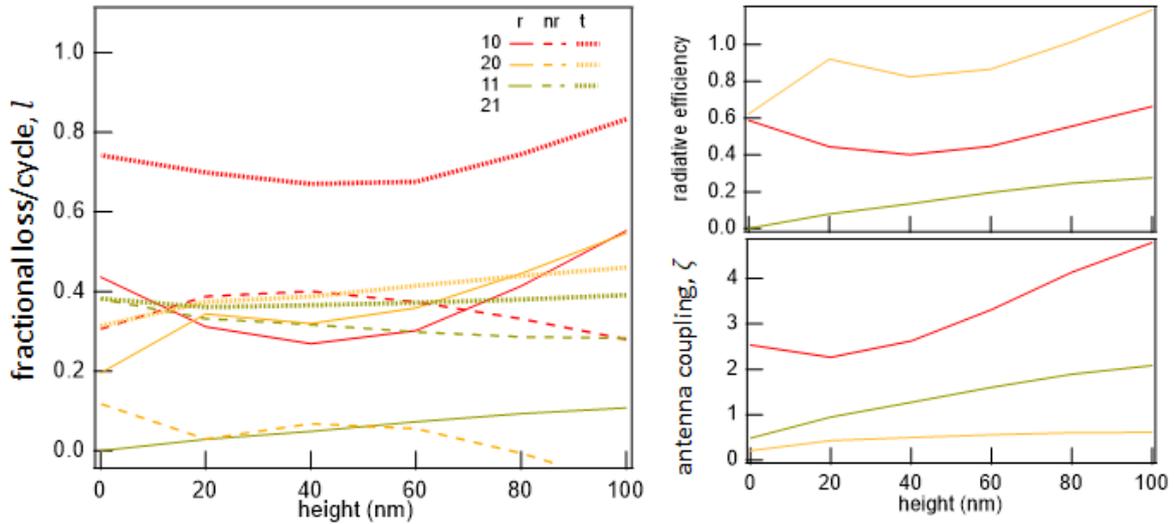

**Figure S5: Scattering properties of NPoM for different dielectric coating heights.** (a) Fractional loss rate per cycle of radiative (r), non-radiative (nr) and total (t) components *vs* dielectric coating height, for each mode. (b,c) Corresponding radiative efficiency and coupling effectiveness (antenna cross section).

This gives an understanding for how the dielectric layer coating acts (Figure S5) to give:
- increased E with dielectric layer height from the increased antenna cross section, strongest for (20, 11) modes,
- strong increase of (11) in near-field and scattering due to increases in both in/out coupling ($\zeta,\eta$),
- (20) radiating strongly, but has poor in-coupling hence only appears weakly in $S$ for low $h$,
- larger linewidth of (20) which comes as both radiative and non-radiative emission is faster,
- a linewidth of (11) dominated by non-radiative components, while (20) is dominated by radiative components.

From this it is clear that the dielectric layer helps in-coupling, but particularly for higher order modes. The dielectric layer also drastically increases the outcoupling for the (11) mode, which can be intuitively considered as helping plasmons escape out around the edge of the nanocavity onto the surface of the NP.

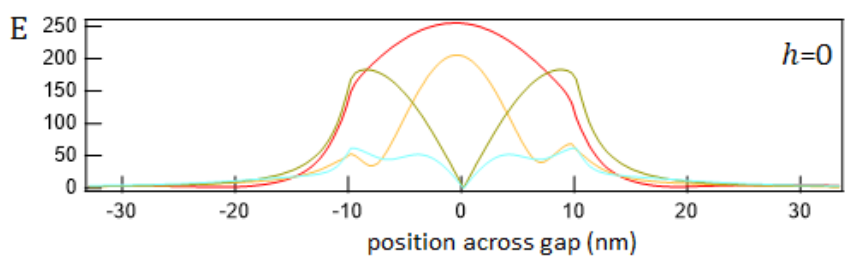

**Figure S6:** Near-field enhancement across the centre of the gap, for no dielectric coating.

Using this model the SERS enhancements can be predicted, since the enhancement (ignoring input/output wavelength differences) is $\eta E^4$.[3] The nanocavity centre favours 'even' modes due to the central antinode, but we integrate over all molecules overlapping the modes to get the predicted SERS emission $= \eta E^4 V/(\sigma^2 d)$ where $\sigma$ is the size of a molecule (Figure S7).

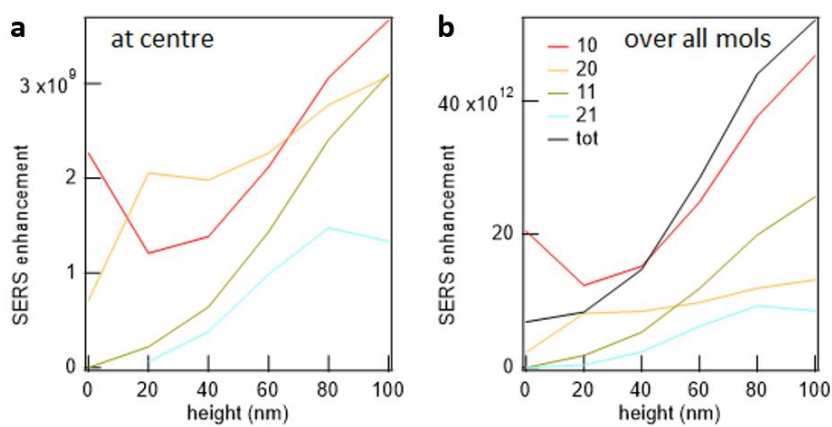

**Figure S7:** SERS enhancement in NPoM as a function of dielectric coating height. a) scaling at centre position, b) and integrated over all molecules.

This model shows that, independent from wavelength tuning, as a result of the retardation matching alone the local refractive index ($n_d$) can provide an increase in SERS of nearly 10x when fully embedded (here calculated for $n_d$=1.6).

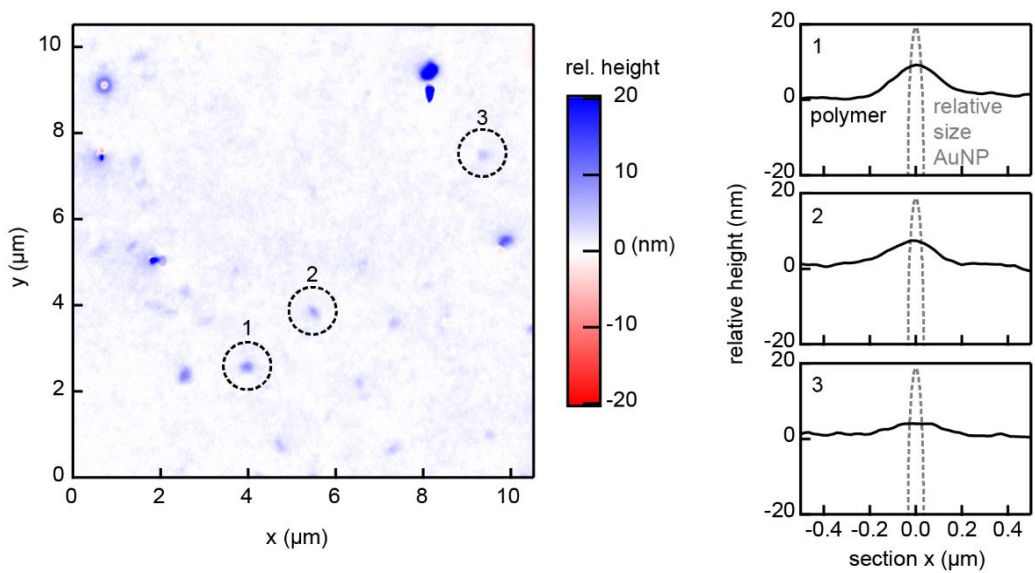

**Figure S8: AFM measurements of polymer coated ($n_d$=1.49) NPoMs** showing residual 2-9 nm local deformations of the surface normal as a result of the underlying nanoparticle inside the embedding film of nominal height 119 nm.